# Quasinormal mode as a foundational framework for all electromagnetic Fano resonances

Mikhail Bochkarev[1], Nikolay Solodovchenko[1], Kirill Samusev[1,2], Mikhail Limonov[1,2], Tong Wu[3], and Philippe Lalanne[3]

[1]School of Physics and Engineering, ITMO University, Saint-Petersburg 197101, Russia
[2]Ioffe Institute, Saint-Petersburg 194021, Russia
[3]Laboratoire Photonique Numérique et Nanosciences (LP2N), Université de Bordeaux, Institut d'Optique Graduate School, CNRS, Talence, France

mikhail.bochkarev@metalab.ifmo.ru
philippe.lalanne@institutoptique.fr

**Abstract.** Fano profiles are observed across various fields of wave physics. They emerge from interference phenomena and are quantified by the asymmetry parameter $q$. In optics, $q$ is usually considered as a phenomenological coefficient obtained by fitting experimental or numerical data. In this work, we introduce an ab initio Maxwellian approach using quasinormal modes to analytically describe line shapes in light scattering problems. We show that the response of each individual quasinormal mode inherently exhibits a Fano profile and derive an explicit analytical formula for the Fano parameter. Experimental and numerical validations confirm the formula's accuracy across a broad spectrum of electromagnetic systems. The general expression for $q$ opens new possibilities for fine-tuning and optimizing spectral line shapes in electromagnetism.

**Introduction.** Interference is a universal phenomenon in wave physics. One of its most notable features is the emergence of asymmetric spectral signatures, arising from the interaction of coherent wave processes through different pathways. This feature was first described by Ugo Fano in 1935 [1] and is now referred to as Fano resonance. Fano resonance has been observed in various branches of physics [2-5], including photonics [6], plasmonics [7], acoustics [8], atomic physics [9], quantum electrodynamics [10], lasers [11], and slow light [12]. In optics, recent developments have unveiled striking examples of this phenomenon, such as bound states in the continuum [13], the Kerker effect [14], and non-radiating anapoles [15].

A conventional way to interpret the asymmetric Fano signatures is to consider the interference between a broad resonance (continuum) and a narrow one (localized state labelled by the subscript $m$), which is described by the Fano profile:

$$\sigma = \sigma_m \frac{(q_m + \Omega_m)^2}{(\Omega_m^2 + 1)(q_m^2 + 1)} = \sigma_m \left[ \frac{q_m^2 - 1 + 2q_m\Omega_m}{(\Omega_m^2 + 1)(q_m^2 + 1)} \right] + \sigma_m \frac{1}{(q_m^2 + 1)}, \qquad (1)$$

where $q_m = -\cot(\delta_m/2)$ is the Fano parameter, $\delta_m/2$ the phase shift between the narrow and broad resonances, and $\sigma_m$ represents the intensity of the narrow resonance. $\Omega_m = 2(\omega - \omega_m)/\gamma_m$ denotes the dimensionless frequency, with $\gamma_m$ and $\omega_m$ representing the resonance damping rate and resonance frequency, respectively. The first expression in Eq. (1) is the traditional form of the Fano line shape, while the second one separates the line shape (first term) from a constant background (second term).

The parameter $q_m$ serves to distinguish asymmetric Fano from symmetric Lorentzian profile. The sign and magnitude of $q_m$ describe the relative positions of the broad and narrow resonances in the spectrum as well as the nature of their interaction. The Fano profile can manifest either asymmetric or symmetric line shapes, depending on the phase shift $δ_m/2$ between the resonances. When $δ_m/2 = 0$, the Fano profile simplifies to a Lorentzian shape, $1/(\Omega_m^2 + 1)$, with $q_m \to \pm\infty$, indicating constructive interference. Conversely, when $\delta_m/2 = \pm\pi/2$, $q_m = 0$, and the Fano profile adopts a quasi-Lorentzian form, $-1/(\Omega_m^2 + 1)$, representing destructive interference. Pronounced asymmetry in the spectral line shape emerges when $0 < |q_m| \leq 10$. Additionally, a true zero in the spectrum appears when a narrow resonance interacts with a single continuum.

Ugo Fano's seminal articles [1-2] on interpreting spectra with asymmetric profiles have generated and continue to inspire significant interest across numerous disciplines. However, Fano's theoretical framework has its limitations. It is well-suited for analyzing isolated, sharp spectral features, where clear quantitative insights into underlying mechanisms can be extracted from spectral data. While substantial progress has been made in atomic photoionization using complex spectral analysis [16,17], the application of Fano's theory in optics remains largely phenomenological, lacking a robust electromagnetic foundation and relying heavily on fitted parameters. Developing a general analytical solution for $q_m$ and $\sigma_m$, applicable to arbitrary resonator shapes and materials, would be a major advancement. We have found that quasinormal mode (QNM) theory provides an elegant and comprehensive solution to this challenge.

QNM theory, much like Fano resonances, spreads across various wave disciplines, including mathematics [18], gravitational waves physics [19], quantum mechanics [17], electromagnetism [20]. QNMs have complex frequencies (or energies), $\widetilde{\omega}_m = \omega_m + \frac{i\gamma_m}{2}$, where the imaginary part encodes decay into a continuum. In electromagnetism, this decay into the continuum can originate from a variety of mechanisms, including material absorption and leakage into different channels such as plane waves in free space, waveguide modes in layered media, or "guided" surface plasmons. This diversity adds complexity to the problem and is likely why the Fano approach has, until now, remained predominantly phenomenological. Hereafter, we adopt an ab initio Maxwellian approach utilizing QNM theory [21-22] to derive an analytical expression for both $q_m$ and $\sigma_m$ in the most general form.

Although the connection between electromagnetic QNMs and the Fano response may seem intuitive, it remains underexplored. Previous studies have been limited to a few specific cases. For example, in [23], the density-matrix formalism of two-level quantum systems is combined with QNM theory to predict that the near-field response of a quantum emitter coupled to surface plasmons exhibits Fano line shapes that vary with the spatial coordinate. Similar conclusions have been reached in related theoretical studies on emitter-plasmon systems using slightly different approaches [24-25]. Furthermore, it has been demonstrated that the Purcell-effect spectrum of emitters coupled to spectrally overlapping resonances can also exhibit Fano line shapes [26], a phenomenon well understood within the framework of non-Hermitian complex mode volumes [27]. Across these studies, the Fano line shape consistently arises in the near-field due to the interaction between spectrally overlapping resonances with different quality factors.

In this study, we take a more fundamental approach, examining the general case of electromagnetic resonances excited by arbitrary far-field waves. The approach reveals a profound analogy with quantum mechanics, where Fano line shapes first appeared in the resonance scattering spectra of quasi-bound states in the continuum [2]. Our main finding is that each individual QNM contributes an intrinsic Fano line shape to the extinction spectrum, with parameters that can be described with a close-form expression – without the need for spectral overlap or near-field backaction. This result demonstrates that QNMs naturally embody the interaction between localized resonances and the continuum, making them a particularly convenient theoretical representation of bound states in the continuum. The Fano parameter differs across modes, as each QNM couples uniquely to different decay channels, and as a result, the total extinction spectrum becomes a sum of multiple Fano contributions, each with distinct parameters.

Our analysis is remarkably general, utilizing a QNM regularization formalism based on complex mappings with perfectly matched layers [20,27-29]. This approach remains robust even in the presence of singularities beyond the QNM spectrum, such as branch cuts. It enables us to tackle complex geometries relevant to nanophotonics, including resonators made of dispersive materials on layered substrates. The framework applies broadly to various decay channels, including uniform semi-infinite spaces, surface plasmons, and guided modes in planar or 2D waveguides. Additionally, it accommodates the critical scenario of spectrally overlapping resonances, even in the presence of exceptional points. Consequently, we our Fano theory is has highly general and versatile scope.

**Theory**. Electromagnetic QNMs are source-free solutions of Maxwell equations, $\nabla \times \tilde{\mathbf{E}}_m = -i\tilde{\omega}_m \mu_0 \tilde{\mathbf{H}}_m$, $\nabla \times \tilde{\mathbf{H}}_m = i\tilde{\omega}_m \varepsilon_0 \varepsilon(\tilde{\omega}) \tilde{\mathbf{E}}_m$, which satisfy the outgoing-wave condition for $|\mathbf{r}| \to \infty$ [20]. Hereafter, $\tilde{\mathbf{E}}_m$ and $\tilde{\mathbf{H}}_m$ respectively denote the normalized electric and magnetic fields of the $m^{th}$ QNM, $\tilde{\omega}_m = \omega_m - i\gamma_m$ is its complex eigenfrequency, $\varepsilon_0 \varepsilon$ the relative permittivity (which can be considered as a dispersive tensor), and $\mu_0$ is the permeability of vacuum. The QNM field exponentially decays in time and $\text{Im}(\tilde{\omega}_m) < 0$ with the $\exp(-i\omega t)$ notation. Hereafter, we consider non-magnetic materials. Extending the analysis to magnetic cases is straightforward. Our only assumption is that the materials are reciprocal.

The resonator is illuminated by a monochromatic wave with a frequency $\omega$. The incident wave is referred to as the background field [20] in the presence of a substrate with an electric field denoted by $\mathbf{E}_b(\mathbf{r}, \omega)$. The extinction cross section is given by

$$\sigma_{ext} = \frac{\omega}{2P_0} \int_{V_r} \text{Im}\{\Delta\varepsilon \mathbf{E}_b^* \cdot (\mathbf{E}_s + \mathbf{E}_b)\} d^3\mathbf{r}, \tag{2}$$

where $\mathbf{E}_s$ is the field scattered by the resonator, $P_0$ is the time-averaged Poynting vector of the incident wave, $\Delta\varepsilon = \varepsilon_0(\varepsilon - \varepsilon_b)$ is the dielectric contrast between resonator with permittivity $\varepsilon$ and the background media with permittivity $\varepsilon_b$. The integration is performed over the volume of the resonator $V_r$, where $\Delta\varepsilon \neq 0$. We use a QNM expansion to represent the scattered electromagnetic field $[\mathbf{E}_s, \mathbf{H}_s]$

$$[\mathbf{E}_s(\mathbf{r}), \mathbf{H}_s(\mathbf{r})] = \sum_m \alpha_m(\omega) [\tilde{\mathbf{E}}_m(\mathbf{r}), \tilde{\mathbf{H}}_m(\mathbf{r})], \tag{3}$$

where $\alpha_m$ is the excitation coefficient of the $m^{th}$ QNM. Expansions with only QNMs are generally not complete, especially when there is a substrate (branch cuts) [20]. To account for this possibility and remain as general as possible, we use a formalism based on a QNM-regularization with complex mappings implemented with perfectly matched layers (PMLs). In this formalism, the extension of Eq. (3) encompasses a finite set of QNMs that are unaffected by the mapping and an infinite set of numerical eigenmodes often called PML modes. The incorporation of PML modes guaranties completeness over the whole mapped space [20-21] with a unique expression for the excitation coefficients

$$\alpha_m(\omega) = -\frac{\omega}{\omega - \tilde{\omega}_m} \int_{V_r} \Delta\varepsilon \mathbf{E}_b \cdot \tilde{\mathbf{E}}_m d^3\mathbf{r}, \tag{4}$$

which is derived using the Lorentz reciprocity theorem. This uniqueness will guarantee the uniqueness of the expression of the Fano coefficients later.

Substitution of Eqs. (3) and (4) into Eq. (2), allows us to expand the extinction cross-section as the series of QNM contributions, $\sigma_{ext} = \sum_m \sigma_{ext}^m$. This classical step is then followed by a series of algebraic manipulations. These manipulations, taken individually, do not pose intrinsic difficulties. It is rather the sequence that is not trivial. We were indeed helped by our intuition to identify a Fano-like response for every $\sigma_{ext}^m$ when we conducted the manipulations. The derivation is lengthy and technical; it is documented in the Suppl. Section I.

We found that $\sigma_{ext}^m$ can be conveniently expressed as

$$\sigma_{ext}^m = \frac{1}{P_0} \frac{\omega^2}{2\gamma_m} |\xi_m| \left[ \frac{\left(-\frac{Re\{\xi_m\}+|\xi_m|}{Im\{\xi_m\}}\right)^2 - 1 + 2\left(-\frac{Re\{\xi_m\}+|\xi_m|}{Im\{\xi_m\}}\right)\Omega_m}{(\Omega_m^2+1)\left(\left(-\frac{Re\{\xi_m\}+|\xi_m|}{Im\{\xi_m\}}\right)^2 + 1\right)} \right], \tag{5}$$

with the term $\xi_m$, defined as the product of two independent overlap integrals over the resonator volume $V_r$

$$\xi_m \equiv \left(\int_{V_r} \Delta\varepsilon(\omega, \boldsymbol{r}')\tilde{\mathbf{E}}_m(\boldsymbol{r}') \cdot \mathbf{E}_b(\omega, \boldsymbol{r}')d^3\boldsymbol{r}'\right)\left(\int_{V_r} \Delta\varepsilon(\omega, \boldsymbol{r})\tilde{\mathbf{E}}_m(\boldsymbol{r}) \cdot \mathbf{E}_b^*(\omega, \boldsymbol{r})d^3\boldsymbol{r}\right). \quad (6)$$

By identifying with the classical Fano expression $\sigma_m\left[\frac{q_m^2 - 1 + 2q_m\Omega_m}{(\Omega_m^2+1)(q_m^2+1)}\right]$, we obtain the main result of the work: two closed-form expressions for the real-valued Fano parameter $q_m$ and the resonance intensity $\sigma_m$ for every individual QNM

$$q_m = -\frac{Re\{\xi_m\} + |\xi_m|}{Im\{\xi_m\}} = -\cot\left(\frac{\arg(\xi_m)}{2}\right), \quad (7)$$

$$\sigma_m = \frac{1}{P_0}\frac{\omega^2}{2\gamma_m}|\xi_m|. \quad (8)$$

Equations (6)-(8) are derived with great generality, imposing no restrictions on the resonator's shape, material characteristics, or background substrate. Importantly, the derivation is free from approximations, relying only on the assumption of material reciprocity.

However, due to the potential $\omega$-dependence of $\Delta\varepsilon$ and $\mathbf{E}_b$ in Eq. (6), $q_m$ and $\sigma_m$ in Eqs. (7)-(8) are also $\omega$-dependent, preventing the assignment of a unique Fano factor to each resonance. This issue can be easily resolved by assuming that $q_m$ and $\sigma_m$ vary more slowly than the Lorentzian factor $1/(\Omega_m^2 + 1)$. In the following analysis, we therefore neglect the $\omega$-dependence of $q_m$ and $\sigma_m$ and calculate $\xi_m$ at the real part of the eigenfrequency $\xi_m \equiv \xi_m(\omega_m)$. This approximation, referred to as the "resonant frequency approximation" (RFA) hereafter, is robust and highly accurate in general, even for resonances with quality factors as low as 5-10 (Suppl. Section II).

The term $\xi_m$ plays a central in our analysis. A comparison of Eqs. (1) and (7) reveals that $\arg(\xi_m)$ corresponds to the phase shift $\delta_m$ between the narrow and broad resonances in the conventional coupled oscillators models of Fano resonances. However, our approach provides a novel perspective: in traditional coupled oscillator models, the phase shift $\delta_m$ is typically regarded as an intrinsic property of the resonance. In contrast, the present theory demonstrates that $\arg(\xi_m)$ explicitly depends on the incident field. Hereafter, this dependency will be clearly illustrated using experimental data, showing that the Fano parameter can be tuned from 0 to $\pm\infty$ by adjusting the direction of the incident field.

Notably, $\xi_m$ incorporates two overlap integrals, allowing the incident field $\mathbf{E}_b$ and its complex conjugate $\mathbf{E}_b^*$ to be interchanged. This property arises from the fact that conjugating the incident field reverses its propagation direction, as $\mathbf{E}_b^*(\boldsymbol{k}_b) = \mathbf{E}_b(-\boldsymbol{k}_b)$. As a result, $\xi_m$ remains unchanged under a reversal of the incident wave direction, i.e. $\xi(\boldsymbol{k}_b) = \xi(-\boldsymbol{k}_b)$. As a results, $q_m, \sigma_m$ and $\sigma_{ext}^m$ are all invariant. It is well established that, in reciprocal systems, the total extinction remains unchanged regardless of whether the system is illuminated from one side or the opposite side, $\sigma_{ext}(\boldsymbol{k}_b) = \sigma_{ext}(-\boldsymbol{k}_b)$ [30,31]. Here, we extend this principle by demonstrating that the extinction "per resonance" and the associated Fano parameters are also invariant under this directional reversal.

Only two earlier works have proposed expressions for the Fano parameter. In [32], an expression for $q_m$ was derived for the Mie multipoles of a spherical particle in free space using Bessel and Hankel functions. However, since Mie multipoles do not correspond to resonances with complex frequencies, this approach is both more limited in scope and notably different from the present one. The second study is conceptually closer, as it bridges Fano parameters to QNMs [33]. However, its derivation relies on restrictive symmetry assumptions, which hinder the analysis of practical resonator cases on substrates. Moreover, as demonstrated in Suppl. Section III.2, the formula proposed in [33] is inaccurate even when the symmetry conditions hold. Our comparison shows that Eq. (7) offers a significantly better fit. Furthermore, the formula in [33] is not invariant under reversal of the incident wave direction, i.e. $q_m(\boldsymbol{k}_b) \neq q_m(-\boldsymbol{k}_b)$, indicating fundamental flaws in the derivation.

**Experimental verification.** To validate the theory, we consider a split-ring resonator with a rectangular cross section made from a high-index, low-loss ceramic compound, $(Ca_{0.67}La_{0.33})(Al_{0.33}Ti_{0.67})O_3$, with a relative permittivity $\varepsilon = 43+0.0034i$, which remains nearly constant in the 1–8 GHz frequency range used in the experiment. A photograph of the split-ring is shown in Fig. 1(a).

To measure the extinction spectrum, we place the split-ring between two horn antennas positioned 2 meters apart in an anechoic chamber (see details in Suppl. Section IV). The spectrum was inferred by comparing the $S_{21}$ measurements with and without the sample. The broken rotational symmetry of the split-ring offers an ideal way to study the Fano parameter for different orientations φ between the sample and the incident light (Fig. 1(a)). Figure 1(b) displays four extinction spectra measured for φ = 0°, 30°, 60° and 90°, which, consistently with Eqs. (7)-(8), demonstrate that Fano parameters strongly depend on the incident background field.

The resonator supports longitudinal Fabry-Pérot-like resonances, indexed by the number $m$ of half-wavelengths fitting within the resonator optical length. Figure 1(c) shows the near-field distributions recorded at resonance for $m = 30$ (red) and 31 (green), approximately 2 mm above the split-ring, using a magnetic near-field probe. The two modes exhibit different symmetries.

The measured spectra are then fitted using the classical Fano formula (Eq. 1), with an additional background intensity term $\sigma_{bg}$, approximated by a second-order polynomial of the frequency. The red and green curves in Fig. 1(b) illustrate the high accuracy of the fit. The fitted parameters, $q_m$ and $\sigma_m$, are presented in Figs. 1(f, g).

To test the present theory, we compute the QNM eigenfrequencies $\tilde{\omega}_m$ and eigenvectors $\tilde{\mathbf{E}}_m$ with the freeware MAN [22]. The QNMs are normalized using the PML norm [20]. In line with the experiment, the background field is modeled as a linearly polarized plane wave. The theoretical values of $\xi_m$ are then calculated using Eq. (6) using the RFA. They are displayed in Figs. 1(d, e), with no adjustable parameters. Then, we compute the $q_m$ and $\sigma_m$ using Eqs. (7) and (8), see the solid curves in Figs. 1(f, g). The Matlab files and COMSOL models developed for the split-ring and other geometries studied in the Supplementary Materials can be found in the new 'Fano' toolbox of the version 9 of MAN. They are available on Zenodo [22].

A strong quantitative agreement between the fitted values of $q_m$ and $\sigma_m$ (square markers) and the theoretical ones (solid curves) is achieved for both modes over a wide range of $q_m$ and $\sigma_m$ values and for all rotation angles φ. The minor deviation observed around φ = 60° is due to experimental noise and a relatively low value of the resonance intensity for that angle, as shown in Suppl. Section III.1 with numerical frequency-domain simulations of the split-ring extinction.

Overall, the $q$-curves for both modes in Fig. 1(f) exhibit distinctive transitions from $q_m = +\infty$ to $q_m = -\infty$ at specific values of φ, where $\arg(\xi_m)/2 = 0$ (Fig. 1(e)). These dramatic variations in the Fano parameter occur exclusively for high-order modes, where the resonator size exceeds the incident wavelength, causing different parts of the mode to interact with varying phases of the incident wave. Conversely, $|\arg(\xi_m)|/2$ remains close to $\pi/2$ for small angles (φ <15°) for $m = 30$ or for large angles (φ > 75°) for $m = 31$. This behavior is due to vanishing resonance intensity (Fig. 1(g)) and the different symmetries of the two modes.

The agreement between theory and experiment in Figs. 1(f)-(g) provides robust validation of the proposed theoretical framework. Further confirmation is presented in the Supplementary Materials, which include numerical simulations spanning a wide variety of resonator shapes and materials. These simulations encompass diverse scenarios, including dielectric and plasmonic particles in free space and on substrates in Suppl. Sections III.1–4, and additional critical cases, such as low-$Q$ resonators with overlapping (example 1) or coalescing (example 2) resonances at exceptional points in Suppl. Section III.5. Together, these tests affirm the broad applicability of our theory, as expected given the absence of approximations in its derivation.

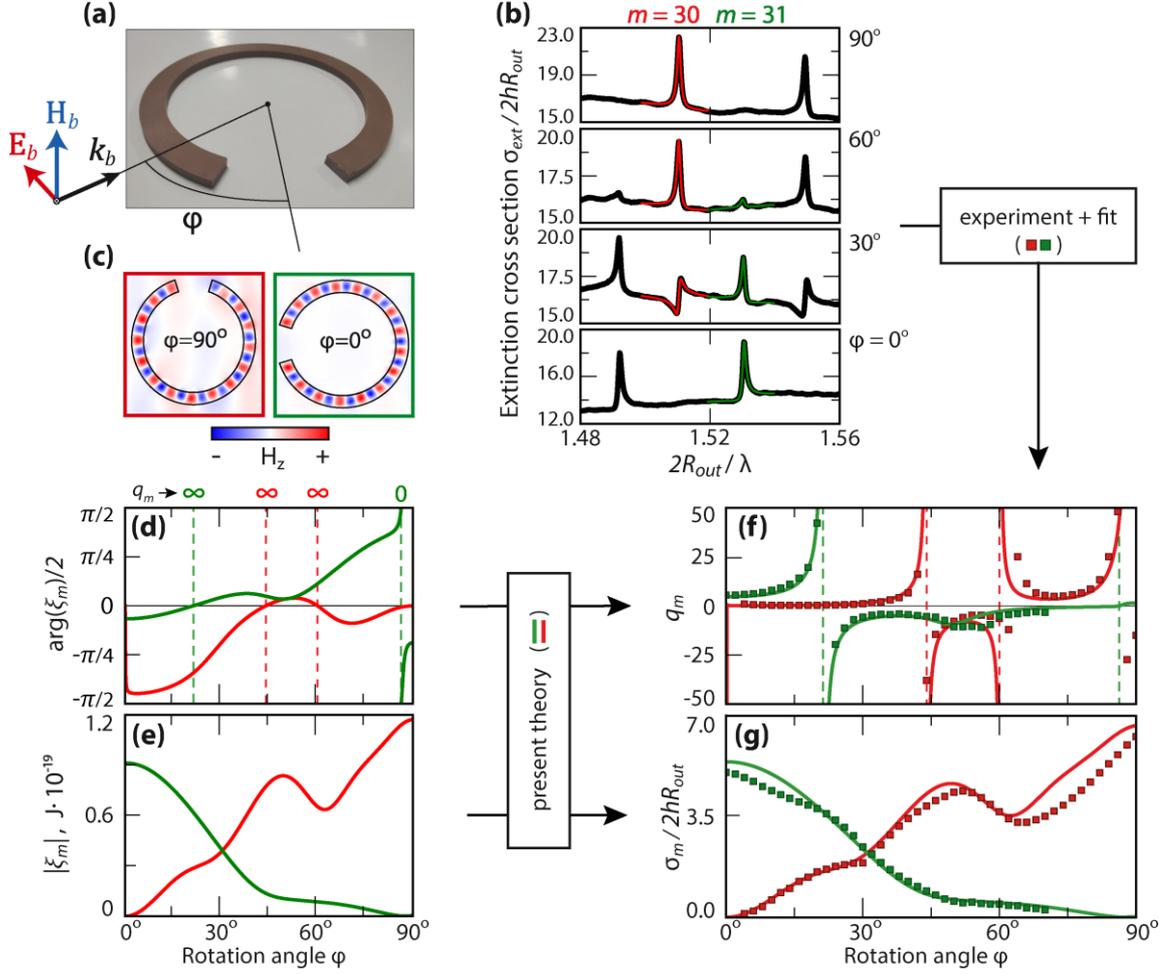

**Fig. 1 – Comparison between the QNM Fano theory and the classical phenomenological approach based on a fit of experimental spectra. (a)** Photograph of the split-ring used in the experiment. **(b)** Measured extinction spectra for several incident angles φ with superimposed green and red lines representing fits using Eq. (1). **(c)** Near-field measurements of the magnetic fields of the anti-symmetric $m = 30$ and symmetric $m = 31$ modes. **(d)** and **(e)** Modulus and argument of $\xi_m$, calculated at the resonant frequency (RFA) using Eq. (6). **(f)** and **(g)** Comparison between theory (Eqs. (7) and (8), RFA) and experiment (Fano fit with Eq. (1)) for the Fano parameter $q_m$ and resonance intensity $\sigma_m$. Vertical dashed lines in (d, f) indicate the theoretical positions of $q_m = 0, \pm\infty$. The resonator parameters are: a height-to-width ratio $h/(R_{out} - R_{in}) = 0.36$, inner and outer radius ratio $R_{in}/R_{out} = 0.81$ with $R_{out} = 57.5$ mm, and a half-gap angular size $\beta = 17.5°$. Note the color labeling consistently used in all panels: $m = 30$ (red) and 31 (green).

**Conclusion.** This work positions QNM theory of scattering as central to the broader family of "Fano effects", which influence many areas of physics, including various contemporary photonic problems with completely different decay physics, such as material absorption or leakage through various radiative channels. It presents QNMs as a unifying framework that integrates discrete and continuum states, by encapsulating critical information about the phases of both narrow and broad resonances all in one. This perspective offers a refreshing alternative to conventional coupled oscillator models of Fano resonances.

Beyond their broad theoretical scope, the expressions for the Fano asymmetry parameter $q_m$ and intensity $\sigma_m$ in Eqs. (7-8) are both simple and intuitive. Calculating these parameters requires only the normalized QNM fields, which are readily accessible through modern QNM solvers. This simplicity enables rapid optimization of resonator geometries and illumination conditions. For instance, as shown in Suppl. Section III.4, this approach can be used to optimize the detection limits of optical resonance sensors by tuning the geometry and adjusting the incident beam direction in a very effective way.

More generally, the introduction of an explicit and general formula for Fano asymmetry parameters offers significant potential for enhancing the optimization of spectral parameters in photonics applications. When combined with computational-bounds approaches [34], QNM formalisms for inverse design [35,36], or deep learning approaches [37-39], this framework could drive the next generation of spectral response design and optimization in photonics.

**Acknowledgments.** MB, NS, KS, and ML acknowledges support from the Russian Science Foundation (project No 23-12-00114). PL acknowledges support from the WHEEL (ANR-22CE24-0012-03) Project.